\documentclass[11pt,preprint]{aastex}
\slugcomment{Submitted to ApJL}

\def\kms{$\rm{km~s}^{-1}$}
\def\etal{et al.}

\shorttitle{First Evidence of EIT Wave and Coronal Moreton Wave}
\shortauthors{Chen \& Wu}

\begin{document}

\title{First Evidence of Coexisting EIT Wave and Coronal Moreton Wave
	from {\em SDO}/AIA Observations}
\author{P. F. Chen and Y. Wu}
\affil{Department of Astronomy, Nanjing University, Nanjing 210093,
	China \\
	Key Lab of Modern Astron. \& Astrophys. (Ministry of
	Education), Nanjing University, China}
\email{chenpf@nju.edu.cn}

\begin{abstract}
``EIT waves" are a globally propagating wavelike phenomenon. They were
often interpreted as a fast-mode magnetoacoustic wave in the corona,
despite various discrepancies between the fast-mode wave model and
observations. To reconcile these discrepancies, we once proposed that
``EIT waves" are apparent propagation of the plasma compression due to
successive stretching of the magnetic field lines pushed by the erupting
flux rope. According to this model, an ``EIT wave" should be preceded by
a fast-mode wave, which however was rarely observed. With the
unprecedented high cadence and sensitivity of the {\em Solar Dynamics
Observatory} ({\em SDO}) observations, we discern a fast-moving wave
front with a speed of 560 km s$^{-1}$, ahead of an ``EIT wave", which
had a velocity of $\sim 190$ \kms, in the ``EIT wave" event on 2010 July
27. The results, suggesting that ``EIT waves" are not fast-mode waves,
confirm the prediction of our fieldline stretching model for ``EIT
wave". In particular, it is found that the coronal Moreton wave was
$\sim 3$ times faster than the ``EIT wave" as predicted.
\end{abstract}


\section{Introduction}\label{sect:intro}

Any perturbation, from small-scale convective motions to large-scale
eruptions, can generate waves propagating in the solar atmosphere.
According to the magnetohydrodynamic (MHD) theory, they can be
identified in terms of fast-mode, Alfv\'enic mode, or slow-mode waves.
For instance, H$\alpha$ Moreton waves were successfully
explained as coronal fast-mode magnetoacoustic waves sweeping the
chromosphere \citep{uchi68}. After the launch of the {\it Solar and
Heliospheric Observatory} ({\it SOHO}) spacecraft, one of its payloads,
EUV Imaging Telescope (EIT), discovered a large-scale wavelike
phenomenon, which was later called ``EIT waves" \citep{mose97, thom98}.

``EIT waves" are bright fronts visible in the EUV difference images,
propagating outward with a typical velocity of 170--350 \kms\
\citep{klas00}. Although the velocity might be underestimated due to the
low cadence of the EIT telescope \citep{long08}, it can really be as
small as $\sim 10$ \kms\ \citep{zhuk09}. In the base difference images,
``EIT waves" are followed by extending dimmings \citep{thom00}. Upon
discovery, ``EIT waves" were soon explained as the fast-mode
magnetoacoustic waves, i.e., the coronal counterparts of H$\alpha$
Moreton waves (or coronal Moreton waves) \citep{thom98}. The ensuing
modelings \citep{wang00, wu01} and observations \citep{vrsn02, warm04,
ball05, grec08, pomo08, vero08, gopa09, pats09} further claimed that the
fast-mode wave model can account for the ``EIT waves". Such a fast-mode
wave model was first questioned by \citet{dela99}, since they
found that an ``EIT wave" stopped at the magnetic separatrix, which is
hard to be understood in the wave model. The second discrepancy between
the fast-mode wave model and observations is that the EIT wave speed
is typically $\sim 3$ times smaller than that of Moreton waves
\citep{klas00}. In addition, it was found that the ``EIT wave" front is
cospatial with the coronal mass ejection (CME) frontal loop
\citep{attr09, chen09a, dai10}, which is not expected from the fast-mode
wave model. In order to reconcile these discrepancies, many efforts have
been taken, and several alternative models have been proposed, e.g., the
fieldline stretching model \citep{chen02, chen05}, the successive
reconnection model \citep{attr07}, the slow-mode wave model
\citep{will06, wang09}, the current sheet model \citep{dela08}, among
others \citep[see][for reviews]{willat09, galllo10, warm10}. 

According to the fieldline stretching model of \citet{chen02, chen05},
a fast-mode wave should be ahead of the ``EIT wave" in a
CME event, providing that the observational cadence and the
sensitivity are high enough. In particular, the model
predicts that the fast-mode wave is $\sim 3$ times faster than the
``EIT wave" if the magnetic configuration is semi-circular. The
coexistence of a faster wave ahead of an ``EIT wave" was discovered by
\citet{harr03}, who revealed with the {\it Transition Region and
Coronal Explorer} ({\em TRACE}) observations that a faint wave front
with a speed of $\sim 500$ \kms\ dispersed out of the
bright ``EIT wave" front, whose speed was only $\sim 200$ \kms. In this
Letter, we report the evidence of a coexisting fast-mode wave
ahead of the slower-moving ``EIT wave" on 2010 July 27, which
was observed by the newly-launched {\it Solar Dynamics Observatory}
({\em SDO}) mission. 

\section{Observations and Data Analysis}
\label{sect:data}

On 2010 July 27, a microflare, with the Solar Object Locator of 
SOL2010-07-27T08:46:00L223C108, occurred to the west side of NOAA
active region (AR) 11089 (S24W21). The microflare was located at S18W48
in the heliocentric coordinates. The GOES 1-8 {\AA} soft X-ray
lightcurve shows a small bump less than A2 level from 08:46 UT to 08:56
UT. Accompanying the occurrence of the tiny flare, EUV waves and
dimmings can be identified in the base difference
images, extending from the flare site to the north for a short
distance. No coronagraph observations from the {\em SOHO} satellite are
available. The microflare was located at S22E31 in the field of view of
the {\em Solar TErrestrial RElations Observatory}
\citep[\textit{STEREO}][]{kais08} A, and the its coronagraphs
did not detect any CME during this event. Two factors do not
favor the detection of a possible CME for this event: one is that the
source region is too close to the solar disk center for the {\em
STEREO} satellite, and the other is that the CME, if existing, should
have a small angular extension as implied by the EUV dimmings, and
therefore is prone to be missed by coronagraphs \citep{zhangu10}.

The microflare and the associated ``EIT wave" were well documented by
the Atmospheric Imaging Assembly (AIA) on the {\em Solar Dynamics
Observatory} \citep[\textit{SDO}][]{titlho06}. The AIA instrument has 10
EUV and UV channels with a spatial resolution of $1.\arcsec 2$. Since
``EIT waves" are most evident at the 193 {\AA} channel, we mainly use
this channel to study the dynamics of the ``EIT wave". The cadence of
the observation is 12 s, which allows the detection of the detailed
features of ``EIT waves" that had not been seen before \citep{liuni10}.

In order to clearly show the propagation of the faint ``EIT waves",
running or base difference technique was often used.
Throughout this paper, we rely on the percentage difference images,
which are obtained by the base difference images divided by the
pre-event image \citep{will99}. The rotation of the Sun is corrected. 
For simplicity, the percentage difference images are mentioned
as difference images or difference intensity hereafter.

\section{Results}
\label{sect:res}

\begin{figure}
\epsscale{1.14}
\plotone{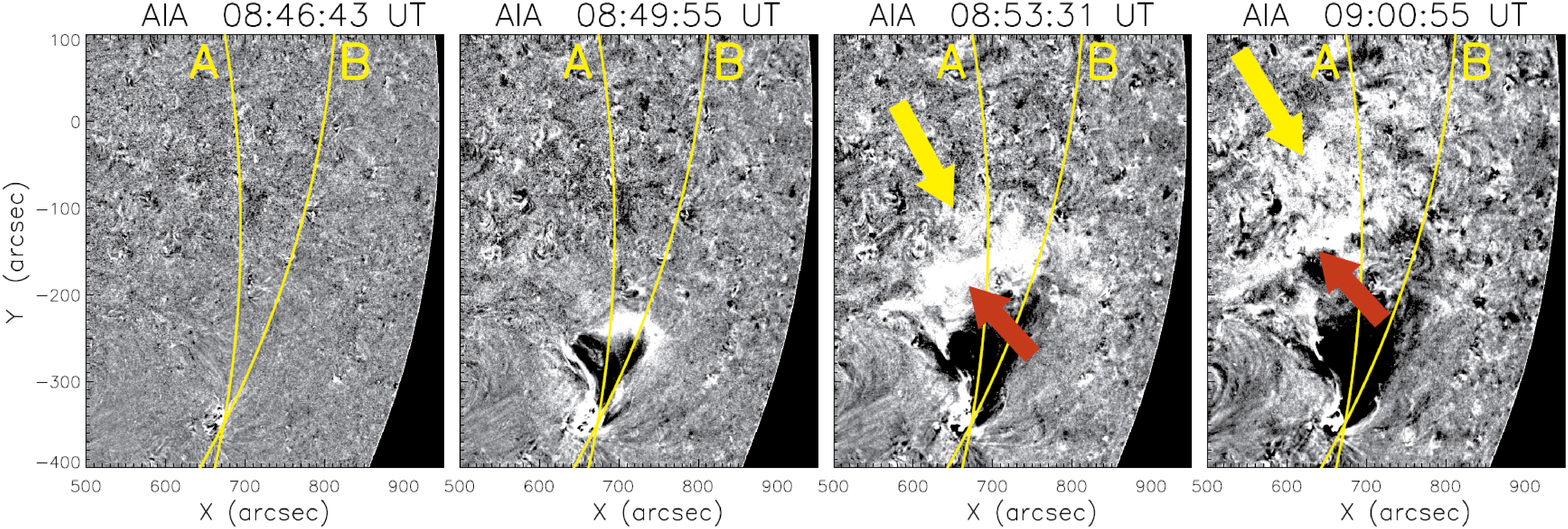}
\caption{Sequential difference images of the 2010 July 7 event
observed by {\em SDO}/AIA at 193 \AA\, where the intensity map at
08:44:55 UT is subtracted. The white/dark pixels indicate intensity
increase/decrease. The red arrows point to the fronts of the slower
``EIT wave", whereas the yellow arrows to the fast-mode wave. Two
slices, A and B, are great circles linking the flare site at 
(670\arcsec, $-350\arcsec$). Note that there exists a small
coronal loop at (720\arcsec, $-200\arcsec$). The corresponding movie,
{\it basediff.mpg}, is attached as online materials.}
\label{fig1}
\end{figure}

The evolution of the ``EIT wave" event on 2010 July 27 is displayed in
Figure \ref{fig1}, which shows the 193 {\AA} difference images at four
moments. The pre-event intensity map at 08:44:55 UT is chosen as the
base image that is subtracted. In this figure, the bright (dark) pixels
indicate the intensity increase (decrease). It is seen that at 08:46:43
UT, a small-sized brightening occurred near (670\arcsec, $-350\arcsec$).
Note that a small coronal loop with relatively stronger magnetic field
was to the north at around (720\arcsec, $-200\arcsec$), which is clearly
seen in the second panel. Shortly later, bright fronts appeared and
began to propagate to the north. It is revealed from the attached movie
({\em basediff.mpg}) that the western and eastern flanks formed first.
At 08:49:55 UT, the northern flank became to be dominant in brightness.
Until this stage, only one bright strip was evident along the northern
flank. At 08:53:31 UT, the northern flank of the brightenings swept over
the small coronal loop, and was then split, presumably due to the small
coronal loop. On the east side, a weak front became detached ahead of
the main strip, as respectively indicated by the yellow and red arrows
in the third panel of Figure \ref{fig1}. The weak front propagated much
faster than the main strip, and the two fronts were widely separated at
09:00:55 UT, as illustrated by the right panel of Figure \ref{fig1}. On
the west side, only a bright strip is discernable by eyes, which
propagated quickly.

In order to investigate the dynamics of the wave fronts more
quantitatively, we analyze the time evolution of the brightness
distributions along two slices, A and B, which are separated by the
small coronal loop at (720\arcsec, $-200\arcsec$) as indicated in Figure
\ref{fig1}. The two slices are both great circles along the solar
surface linking the flare site. The results are displayed in Figure
\ref{fig2} as a space-time plot, where any oblique pattern means a
propagating front.

The top panel of Figure \ref{fig2} displays the evolution of the wave
front propagation along slice A. We can identify several fronts: F1, F2,
and S1. Front F1 appeared first, which was propagating with a high
velocity of 560 \kms\ until 08:52:44 UT. We interpret it as a fast-mode
wave. A separate bright front S1 was propagating with an initial speed
of 190 \kms\ since 08:52:00 UT. Its speed was decreasing. After 09:02:28
UT, front S1 stopped at a distance of $\sim 253\arcsec$ from the flare
site. We identify it as the classical ``EIT wave". Front S1 might be
already existing during the period from 08:49:04 UT to 08:52:00 UT, but
it is almost undistinguishable from the fast front F1 in Figure
\ref{fig2}. It is interesting to see that an additional wave pattern,
front F2, emanated from front S1 at the distance of $\sim 220\arcsec$
from the flare site at 08:54:54 UT. This weak front propagated with a
speed of 310 \kms. We interpret it as a fast-mode wave. An important
reason is that after a distance of $\sim 270\arcsec$ the fast front F1
began to decelerate, and propagated with almost the same speed as front
F2.

The bottom panel of Figure \ref{fig2} illustrates the wave front
propagation along slice B. The wave pattern here is very
simple. Besides a fast-moving bright front, F3, which had a
velocity of 470 \kms, there is a weak front, S2, propagating outward
with a velocity of 170 \kms. We identify the faster front F3 as a
fast-mode wave and the slower front S2 as an ``EIT wave".

\section{Discussions}
\label{sect:Dis}

\subsection{Identification of ``EIT Wave" and Coronal Moreton Wave}

Since the discovery, ``EIT waves" were widely explained as fast-mode
magnetoacoustic waves, or coronal Moreton waves, as reviewed in
\S\ref{sect:intro}. However, many of their characteristics cannot be
explained in the framework of fast-mode waves, which stimulated
\citet{dela00} to doubt the wave nature of the ``EIT waves". Based on
MHD simulations, \citet{chen02, chen05} proposed a magnetic
fieldline stretching model, i.e., ``EIT waves" are not fast-mode waves,
they are apparent propagation of the brightenings from the compressed
plasma generated by successive stretching of the magnetic field lines
overlying the erupting flux rope. The model predicts that there should
be a sharp fast-mode wave, or coronal Moreton wave, which is
piston-driven by the erupting flux rope and propagates ahead of the
``EIT waves". The speed of the faster wave is just the local fast-mode
magnetoacoustic wave speed; However, the ``EIT wave" speed is determined
by both local parameters and the magnetic configuration. According to
this model, if the coronal magnetic field lines are semi-circles, the
``EIT wave" should be $\sim 3$ times slower than the fast-mode wave. If
the coronal field has a strongly stretched configuration, the resulting
``EIT wave" has a very low speed \citep{chen09b, yang10}, and is close
to zero near a magnetic separatrix. Such a fieldline stretching model,
which was originally based on 2-dimensional simulations, was recently
backed by 3-dimensional MHD simulations \citep{down11}.

\citet{chen02, chen05} pointed out that the coronal Moreton waves are so
fast that at most one front can be detected with the low-cadence ($\sim
15$ min) observations of the {\em SOHO}/EIT telescope, e.g., the sharp
front in Fig. 1 of \citet{thom00} is definitely a coronal Moreton wave
front. Their propagation can be caught only with a higher cadence.
However, even with a cadence of 2.5 min, the Extreme Ultraviolet Imager
(EUVI) aboard {\em STEREO} still has not detected a coronal Moreton
wave ahead of any ``EIT wave". Up to now, the existence of a coronal
Moreton wave ahead of the ``EIT wave" was implied by the filament
winking \citep{eto02}, and the only direct evidence was revealed by
\citet{harr03} with the {\em TRACE} observations, though \citet{will06}
studied the same event and claimed that there is no wave front ahead of
the ``EIT wave". According to \S\ref{sect:res}, it is seen that the
high-resolution observations of the {\em SDO}/AIA imaging telescope
convincingly revealed that there is another faster wave propagating
ahead of the slower wave, either along slice A or slice B as marked in
Figure \ref{fig1}. We interpret the slower wave, i.e., fronts S1 and S2
in Figure \ref{fig2}, as ``EIT waves", and the faster waves, i.e.,
fronts F1--F3 in Figure \ref{fig2}, as the fast-mode magnetoacoustic
wave, or coronal Moreton wave.

Along slice A, the top panel of Figure \ref{fig2} clearly shows that
the slower front S1 propagated outward with bright and diffuse fronts
and with an initial speed of 190 \kms, which are the typical
characteristics of ``EIT waves" \citep{klas00}. During the propagation,
the front S1 decelerated, and became almost stationary after
09:02:40 UT, when the front was at a distance of 250\arcsec\ from the
flare site. Previously, the data analysis of \citet{dela99} indicated
that ``EIT waves" stop at the footpoints of a
magnetic separatrix. This feature was successfully explained by
\citet{chen05, chen06} in terms of the fieldline stretching model,
i.e., the magnetic field across the magnetic separatrix belongs to
another flux system, and hence are not involved in the stretching
process. Therefore, ``EIT waves" cannot go across the magnetic
separatrix. To confirm such a conclusion, we plot in Figure \ref{fig3}
the coronal magnetic field near the eruption site, which is
extrapolated based on the {\it SOHO}/Michelson Doppler Imager (MDI)
magnetogram \citep{sche95} with the Potential Field Source-Surface model
\citep{schr03}. The arrow in the figure points to the location where
the ``EIT wave" front stopped. We can see that it is indeed cospatial
with the magnetic separatrix.

The coronal Moreton wave, front F1 in the top panel of Figure
\ref{fig2}, which is fast-mode wave in nature, does not care the
existence of the magnetic separatrix at the distance of 250\arcsec\ from
the flare site. It propagated across it, though its speed decreased
(due to weaker magnetic field) as inferred from its declined slope in
the top panel of Figure \ref{fig2}. However, one feature that is not
expected is the second fast wave front F2 in the top panel of Figure
\ref{fig2}. Apparently it emanated from the bright ``EIT wave" front at
08:55:00 UT when the ``EIT wave" front was at a distance of $\sim 
220\arcsec$ from the flare site. Checking the EUV images in Figure
\ref{fig1}, we find that there is a small-sized coronal loop situated
at such a distance, with a relatively stronger magnetic field as
mentioned in \S\ref{sect:res}. Therefore, we hypothesize that front F2
is due to the fast-mode magnetoacoustic wave near the west side of the
small coronal loop being diffracted to propagate to the east. This is
reinforced by the movie ({\it basedif.mpg}) attached with Figure
\ref{fig1}.

Along slice B, the bottom panel of Figure \ref{fig2} presents a bright
front (F3) with a propagation speed of 470 \kms\ and another faint front
behind (S2) with a speed of 170 \kms. Apparently the faster wave seems
to be an ordinary ``EIT wave" in the sense that an expanding dimming was
immediately following. However, we interpret it as the coronal Moreton
wave, i.e., a fast-mode wave, because it was moving together with the
coronal Moreton wave on slice A. Besides, there is a very faint
reflected wave in the bottom panel of Figure \ref{fig2} when the faster
wave F2 approached the small coronal loop at 08:52:30 UT, which is a
typical characteristic of fast-mode waves. Moreover, this faster wave
has fronts sharper than ``EIT waves". Telling the nature of the slower
front S2 is not straightforward since it was so weak that it is even not
discernable from the difference images in Figure \ref{fig1}. Since this
front had a typical speed for ``EIT waves", we tentatively
explain it as an EIT wave. It is seen that the EIT wave front along
slice B was much weaker than that along slice A. The possible reason is
that the magnetic field lines overlying the eruption site were mainly
oriented along the direction of slice A as illustrated by Figure
\ref{fig3}. According to the fieldline stretching model of
\citet{chen02, chen05}, as these field lines are pushed to stretch up,
compression would be formed at the legs of these field lines.
Therefore, bright ``EIT wave" fronts are visible along slice A. The
fast-mode wave, however, is always refracted toward the region with weak
magnetic field \citep{uchi68}, and has nothing to do with the magnetic
connectivity. This is why the coronal Moreton wave is extremely bright
along slice B (where the magnetic field is weak) and faint along slice
A.

It is interesting to note that, either along slice A or slice B, the
coronal Moreton wave was moving with a speed $\sim 3$ times higher than
that of the following ``EIT wave", consistent with
the prediction of the fieldline stretching model of \citet{chen02,
chen05} when semi-circular magnetic configuration is assumed.

\subsection{Why Were Fast-mode Waves Missed Before?}

\citet{chen02, chen05} predicted that the 
fast-mode wave ahead of the ``EIT wave" was missed by {\em SOHO}/EIT due
to its low cadence of $\sim 15$ min, and would be observed by
imaging telescopes with a higher cadence. However, after the launch of
the {\em STEREO} mission, the EUVI instrument with a high cadence of
2.5 min still did not catch the fast-mode wave \citep{willat09}.

\begin{figure}
\epsscale{0.8}
\plotone{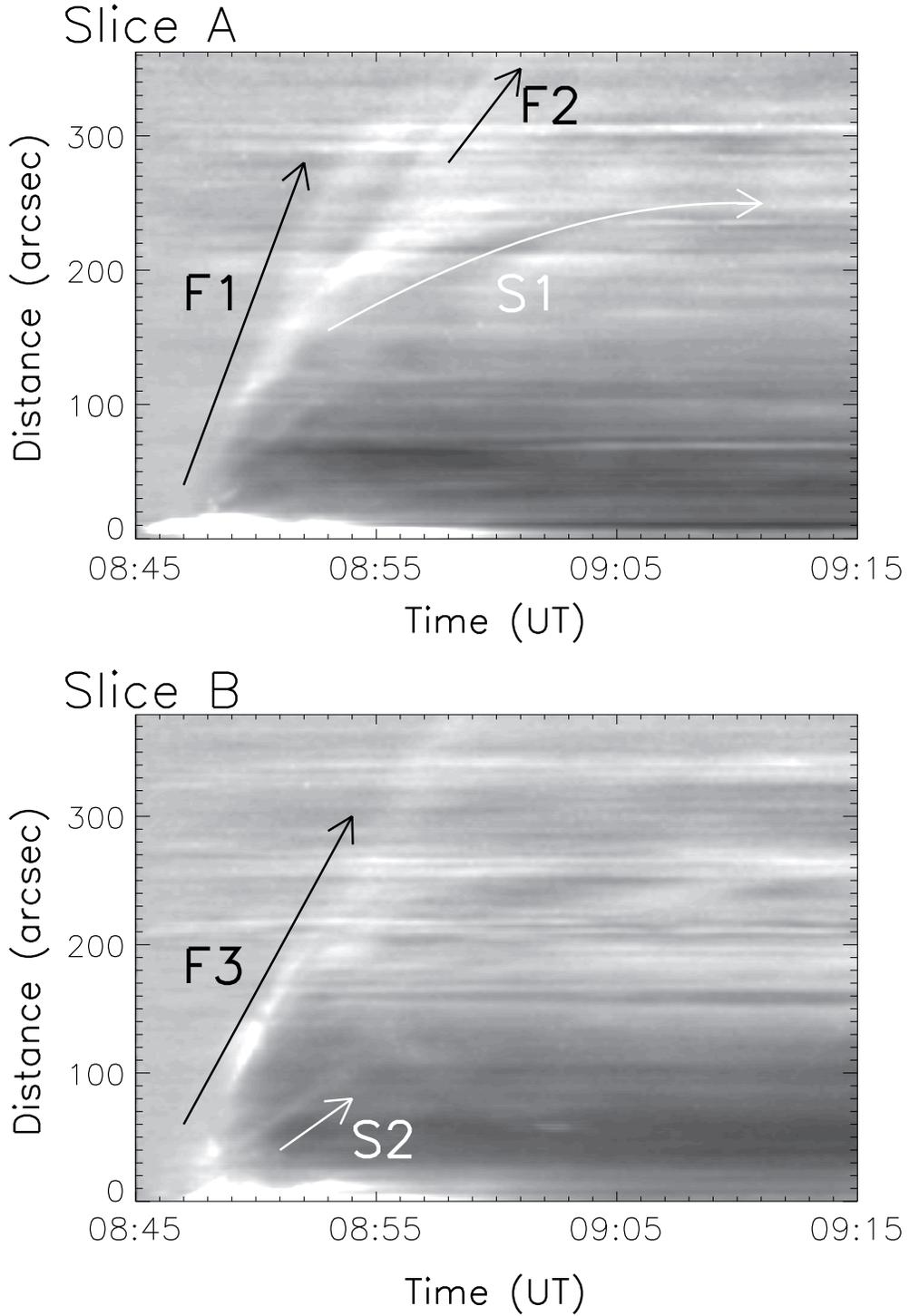}
\caption{Time evolutions of the 193 \AA\ difference intensity
distributions along slice A ({\it top}) and along slice B ({\it
bottom}). The trajectories of slices A and B are marked in Figure
\ref{fig1}, and the distance is measured along the slice from the flare
site. Wave patterns are indicated by the arrows, with the
corresponding speeds of F1 $\sim 560$ \kms, F2 $\sim 310$ \kms, S1
$\sim <190$ \kms, F3 $\sim 470$ \kms, and S2 $\sim 170$ \kms.}
\label{fig2}
\end{figure}

\begin{figure}
\epsscale{1.0}
\plotone{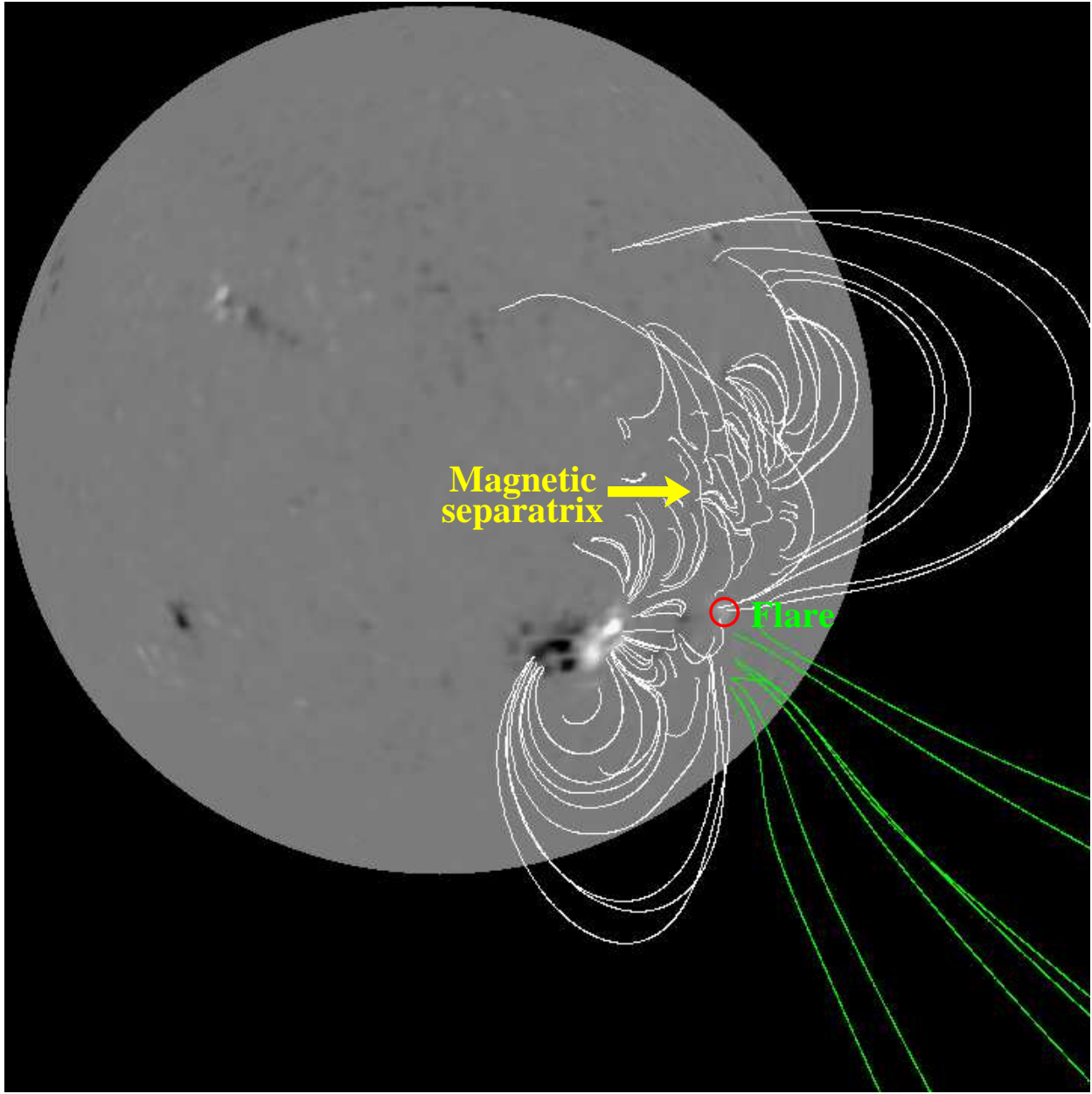}
\caption{Coronal magnetic field distribution near the eruption site,
which is extrapolated from the {\it SOHO}/MDI magnetogram with the
Potential Field Source-Surface model. The yellow arrow points to a
magnetic separatrix.}
\label{fig3}
\end{figure}

\begin{figure}
\epsscale{1.0}
\plotone{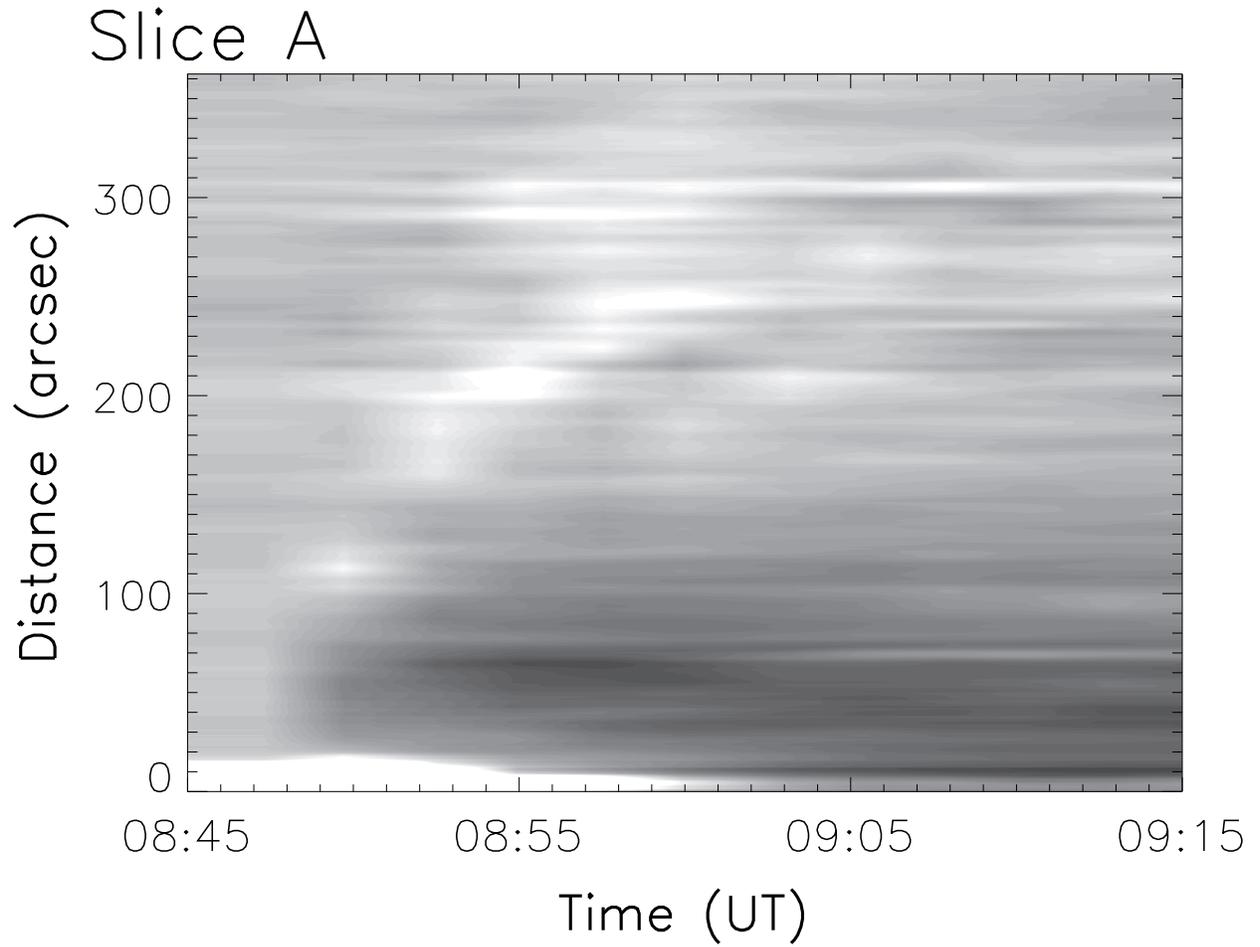}
\caption{Time evolution of the 193 \AA\ difference intensity
distribution along slice A, which is a reproduction of the top panel of
Figure \ref{fig2} but with a degraded cadence of 2.5 min.
}
\label{fig4}
\end{figure}

To check the eligibility of {\em STEREO}/EUVI in detecting the possible
fast-mode wave ahead of the ``EIT wave", we degrade the temporal
resolution of the top panel of Figure \ref{fig2} from 12 s to 2.5 min,
and re-plot it in Figure \ref{fig4}. It is seen that the wave pattern
becomes messy, and we cannot distinguish a faster wave from the ``EIT
wave". This means that, for an ``EIT wave" event like that on 2010 July
27, the 2.5-min cadence of EUV imaging observations even with a spatial
resolution of $1\arcsec .2$ cannot detect the fast-mode wave (or coronal
Moreton wave) ahead of the ``EIT wave". We change the cadence and find
that only if the observational cadence is shorter than 70 s can the
fast-mode wave be distinguished from the ``EIT wave" for the 2010 July
27 event.

With {\em STEREO}/EUVI observations, \citet{cohe09} did show a weak
fast-mode wave component besides the ordinary ``EIT wave". They found
that the fast-mode wave was coupled with the ``EIT wave" when the CME
was expanding laterally, and the two waves ultimately decoupled when the
``EIT wave" front became stationary. Such a result may result from the
low cadence of the {\em STEREO}/EUVI data, as people can get the same
impression from Figure \ref{fig4}. With a high cadence of 12 s, Figure
\ref{fig2} clearly reveals that the fast-mode wave was already distinct
from the ``EIT wave" front before the ``EIT wave" stopped.

To summarize, we analyzed the ``EIT wave" event on 2010 July 27 with the
{\em SDO}/AIA data. It is seen that even for a tiny flare,
the high-resolution observations of the {\em SDO}/AIA telescope still
revealed many wave patterns. In this Letter, we found that a fast-mode
magnetoacoustic wave was propagating ahead of the ``EIT wave". As
predicted by the fieldline stretching model of \citet{chen02, chen05},
the fast-mode wave had a speed $\sim 3$ times higher than that of the
``EIT wave". The fast-mode wave kept propagating after the ``EIT wave"
stopped at the magnetic separatrix. Our results are strongly suggestive
of that ``EIT waves" are not fast-mode magnetoacoustic waves, and can be
well explained by our fieldline stretching model.

\acknowledgments
The authors thank A. Title and N. Nitta for discussions and the AIA
team for providing the calibrated data. {\it SOHO} is a project of
international cooperation between ESA and NASA. This research is
supported by the Chinese foundations 2011CB811402 and NSFC (10403003,
10933003, and 10673004).

\clearpage

\clearpage

\clearpage

\clearpage

\end{document}